\documentclass[aps,pra,twocolumn,showpacs,amsmath,amssymb,superscriptaddress]{revtex4-2}

\usepackage{graphicx}
\usepackage{color}
\usepackage{epstopdf}
\usepackage{listings}
\usepackage{braket}
\usepackage{float}
\usepackage{hyperref}

\begin{document}
\title{Optical Thouless pumping transport and nonlinear switching in a topological low-dimensional discrete nematic liquid crystal array}

\author{Pawel S. Jung}
\thanks{Corresponding author: pawel.jung@ucf.edu}
\affiliation{CREOL, The College of Optics and Photonics, University of Central Florida, Orlando, Florida 32816, USA}
\affiliation{Faculty of Physics, Warsaw University of Technology, Warsaw, Poland}
\author{Midya Parto}
\affiliation{CREOL, The College of Optics and Photonics, University of Central Florida, Orlando, Florida 32816, USA}
\author{Georgios G. Pyrialakos}
\affiliation{CREOL, The College of Optics and Photonics, University of Central Florida, Orlando, Florida 32816, USA}
\author{Hadiseh Nasari}
\affiliation{Ming Hsieh Department of Electrical and Computer Engineering, University of Southern California, Los Angeles, California 90089, USA}
\author{Katarzyna Rutkowska}
\affiliation{Faculty of Physics, Warsaw University of Technology, Warsaw, Poland}
\author{Marek Trippenbach}
\affiliation{Faculty of Physics, University of Warsaw, Warsaw, Poland}
\author{Mercedeh Khajavikhan}
\affiliation{CREOL, The College of Optics and Photonics, University of Central Florida, Orlando, Florida 32816, USA}
\affiliation{Ming Hsieh Department of Electrical and Computer Engineering, University of Southern California, Los Angeles, California 90089, USA}
\affiliation{Department of Physics \& Astronomy, Dornsife College of Letters, Arts, \& Sciences, University of Southern California, Los Angeles, CA 90089, USA}
\author{Wieslaw Krolikowski}
\affiliation{Laser Physics Centre, Research School of
Physics and Engineering, Australian National University, Canberra, ACT
0200, Australia}
\affiliation{Science Program, Texas A\&M University at Qatar, Doha, Qatar}
\author{Demetrios N. Christodoulides}
\affiliation{CREOL, The College of Optics and Photonics, University of Central Florida, Orlando, Florida 32816, USA}

\date{\today}

\begin{abstract}
We theoretically investigate a Thouless pumping scheme in the 1D topological Su-Schrieffer-Heeger (SSH) model for single and multiple band-gaps systems when implemented in a discrete nematic liquid crystal arrangement. For an electrically controlled SSH waveguide array, we numerically demonstrate edge-to-edge light transport at low power levels. On the other hand, at higher powers,  the transport is frustrated by light-induced nonlinear defect states, giving rise to robust all-optical switching.
\end{abstract}

\maketitle


Topologically protected transport is among the principal hallmarks of topological insulators (TIs)\cite{haldane1988model,kane2005quantum,konig2007quantum,Xu2015,Lv2015,jotzu2014experimental}, materials that exhibit an insulating behavior in their bulk, while allowing electron conduction via mid-gap topologically protected edge states on the surface. This protection is naturally manifested in photonic two-dimensional TIs in the presence of structural imperfections or disorder where the chiral edge modes are immune to back-scattering \cite{wang2009observation,liu2021gain, khanikaev2013photonic, hafezi2013imaging,rechtsman2013photonic, bandres2018topological}, thus ensuring a one-way light transport. On the other hand, in one-dimensional topological settings, like that associated with the Su-Schrieffer-Heeger (SSH) model \cite{su1979solitons}, the edge states are zero-dimensional and therefore exhibit no transport properties. However, by using topological pumping schemes (in either space or time) one may be able to  artificially increase the dimensionality of a given  topological configuration. For example, Thouless pumping protocols ~\cite{thouless1983quantization}
can provide a route for edge to edge light transport even in one-dimensional systems \cite{kraus2012topological,verbin2015topological}, which happens to be robust against  random defects.  Within the context of photonics, such Thoulesse schemes have been recently demonstrated in 1D waveguide arrays  ~\cite{cerjan2020thouless,fu2021nonlinear,Rechtsman:nature:21}. Furthermore, the concept of Non-Abelian Thouless pumping in a photonic lattice has also been lately proposed \cite{PhysRevA.103.063518}. 

The functionality of topological systems can be greatly enhanced by utilizing nonlinear interactions. Liquid Crystals (LCs), are known for their strong nonlinearities and thus can be of use in topological photonics ~\cite{LCT}. In Nematic Liquid Crystals (NLCs), the nonlinearity results from molecular reorientation and can be externally tuned via electric \cite{peccianti2000electrically} or magnetic fields \cite{izdebskaya2018stable}.These attributes, 
 \begin{figure}[H]
\centerline{\includegraphics[scale=0.95]{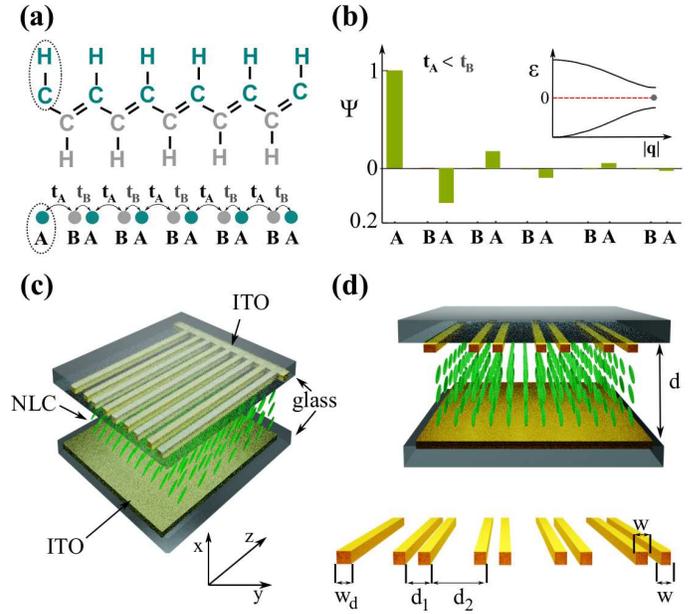}}%
\caption{\label{fig_1} The schematic of the Su-Schrieffer-Heeger model for trans-polyacetylene host and its potential implementation in a discrete liquid-crystal platform. (a) Chemical form showing the two-site unit cell structure (A and B) with the carbon defect marked dashed circle. The tight-binding model representation of hopping electrons with strength $t_{A}$ and $t_{B}$ is shown at the bottom. (b) The bar plot on the right represents the wave function $\psi$ for the edge state with the inset depicting the full band structure. The planar liquid-crystal cell with ITO electrodes as a platform for the SSH model where $d$ represents the cell thickness, $w$ and $w_d$ the electrode thickness and $d_1$, $d_2$ the distances between electrodes is shown in (c) and (d).}
\end{figure} 

\noindent combined with topological configurations can open new vistas 
for novel low power all-optical steering elements that are robust to 
perturbations~\cite{Kivshar,xia2021topological}.

In this work, we theoretically investigate the response of an optical, liquid crystal based topological SSH model in the presence of adiabatic Thouless pumping. In this respect, we provide a design involving a periodic electrode pattern in the NLC planar cell in order to emulate the SSH array that is known to support topologically protected states. In addition, we introduce a $z$-dependent electrode design in order to implement the Thouless pumping scheme. Finally, by using nonlinearity as another degree of freedom, we break the chiral symmetry of the system, thus allowing light switching between topological and trivial defect states.

We start with a brief discussion of the simple 1D non-trivial SSH polyacetylene model, as schematically shown in Fig.1(a)(b). This model is known to describe particle hopping on a one-dimensional chain, with staggered hopping amplitudes ($t_A$ and $t_B$). This finite chain possesses two species A and B, spanning in total M sites. Within the tight-binding formalism, the dynamics in this SSH configuration can be described by the following Hamiltonian \cite{parto2018edge}: 
\begin{eqnarray}
     \widehat{H}=M_{A}\sum_{n=1}^{M} \widehat{c}_{n}^{A\dagger}\widehat{c}_{n}^{A}+M_{B}\sum_{n=1}^{M} \widehat{c}_{n}^{B\dagger}\widehat{c}_{n}^{B} \\ \nonumber +\sum_{n=1}^{M}\left(\widehat{c}_{n}^{B\dagger}\widehat{c}_{n}^{A}+\widehat{c}_{n}^{A\dagger}\widehat{c}_{n}^{B}\right)t_{A}+\left(\widehat{c}_{n-1}^{B\dagger}\widehat{c}_{n}^{A}+\widehat{c}_{n}^{A\dagger}\widehat{c}_{n-1}^{B}\right)t_{B}
\end{eqnarray}
where $\widehat{c}_{n}^{A}\dagger$, $\widehat{c}_{n}^{B}\dagger$ and $\widehat{c}_{n}^{A}$, $\widehat{c}_{n}^{B}$ 
denote creation and annihilation operators at site $n$ in the sublattices A and B, respectively, while $M_{A}$ and $M_{B}$ represents on-site energy offsets. This model can be readily implemented in photonics using a 1D waveguide lattice with two different nearest neighbor coupling coefficients $\kappa_1$ and $\kappa_2$. From coupled mode theory, one finds that the corresponding evolution dynamics in this array are given by:
\begin{equation}
i\frac{d}{dz}\psi_{n}+\beta_{n}\psi_{n}+\kappa_1\psi_{n-1}+\kappa_2\psi_{n+1}=0   
\end{equation}
where $\psi_{n}$ represents the modal field amplitude and $\beta_{n}$ is the propagation constant for waveguide site $n$. Here for convenience we will assume that all sites involved are characterized by the same propagation constant. The same SSH model can be directly realized using a nematic liquid crystal platform. For this case, we consider  the  NLC cell depicted in Fig.1(c), where an AC electric field is imposed externally through the patterned electrodes  in order to induce molecular orientation, that will in turn create a periodic waveguide system. The cell consists of a thin film of nematic liquid crystal of thickness $d$ placed between two glass plates, which provide planar anchoring for molecules in the direction of light propagation. The electrodes in the lower and upper plates are made of  Indium Tin Oxide (ITO). In contrast to previous works where discrete light propagation in LCs was studied ~\cite{fratalocchi2004_OL,fratalocchi2005_OE}, here the upper  electrode pattern features a double spacing  with two separation distances  $d_1$ and $d_2$. Furthermore, these separations can also vary in the $z$-direction. Consequently, the electric field induced waveguides will follow the same pattern. In our configuration, all but one electrodes have the same widths $w$, as it is shown in Fig.1(c). The  width of the  left edge electrode chosen such that  $w_d > w$, ensures that all local modes of each wave-guide should share approximately the same propagation constant. 

The  orientation of the molecules in the LC cell is determined by the externally applied AC field and by the electric field of the optical beam itself. For an extraordinary polarized optical beam, the molecular orientation angle $\theta$ can be obtained from the following equation ~\cite{assanto2003spatial}:
\begin{equation}
    \triangle \theta + \frac{\sin\left(2\theta\right)}{2K}\left( \Delta\varepsilon_{lf}|E_{x}|^{2} + \frac{\Delta\varepsilon}{2}|E|^{2} \right)=0
    \label{Frank}
\end{equation}
where $K$ is the effective elastic constant \cite{khoo1993optically}, while $\Delta \epsilon_{lf} $ and  $\Delta \epsilon $ stand for the low and high frequency dielectric anisotropy, respectively. On the other hand, $E$ represents the $x$ component of the electric field associated with extraordinary polarized beam propagating in the LC cell while $E_{x}$ is the low-frequency field from the biased electrodes. The $E_x$ field can be determined from the corresponding potential $V$ after solving Laplace's equation in this anisotropic medium:
\begin{equation}
    \sum_{ii=x,y,z}\frac{\partial}{\partial ii}\left(\varepsilon_{ii}\frac{\partial}{\partial ii} V\right) = 0,
\end{equation}
 where $\varepsilon_{ii}$ denotes diagonal components of the electric permittivity tensor \cite{Sala:12}. In the absence of linear absorption effects, the evolution of the optical beam (propagating predominantly along the $z$-axis), is described by the following wave equation \cite{fleck1983beam,PhysRevA.82.023806,jung2021formation}: 

\begin{eqnarray}\label{nls}
 && 2ik_{o}n(\theta_{o})\left( \frac{\partial E}{\partial z} + \tan\delta(\theta)\frac{\partial E}{\partial x} \right)+D_{x}\frac{\partial^{2} E}{\partial x^{2}}
+ \left(\theta\right)\frac{\partial^{2} E}{\partial y^{2}} \nonumber \\& &+k_{o}^{2}\left( n^{2}(\theta)-n^{2}(\theta_{o}) \right)E=0,
\end{eqnarray}
\noindent where $ k_o$=2$\pi/\lambda_0$,  $\delta\left(\theta\right)$ is the walk-off angle along the beam axis, $\theta_0$ is the initial 
 molecular orientation (in the absence of light), $D_x=\cos^{2}\theta+\gamma^2\sin^{2}\theta$ is the diffraction coefficient across $z$  and 
$n\left(\theta\right) = \left(\cos^{2}\theta/n_{o}^{2}+\sin^{2}\theta/n^{2}_{e}\right)^{-1/2} $
is an effective index of refraction for the $x$-polarized (i.e., extraordinary) light beam. Here, $n_0$ and $n_e$ represents ordinary and extraordinary refractive indices, respectively and $\gamma^2=n_e^2/n_o^2$.
Assumption of the single elastic constant in Eq.(\ref{Frank}) and restriction of the director movement to the single plane greatly simplifies the formal description of the response of the liquid crystal. We checked that it nevertheless gives results that agree very well with the full vectorial model involving all elastic constants ($K_{11}$,$K_{22}$,$K_{33}$) provided $K=\frac{K_{11}+K_{33}}{2}$ \cite{Sala:12}.

\begin{figure*}[t]
\centerline{\includegraphics[scale=0.95]{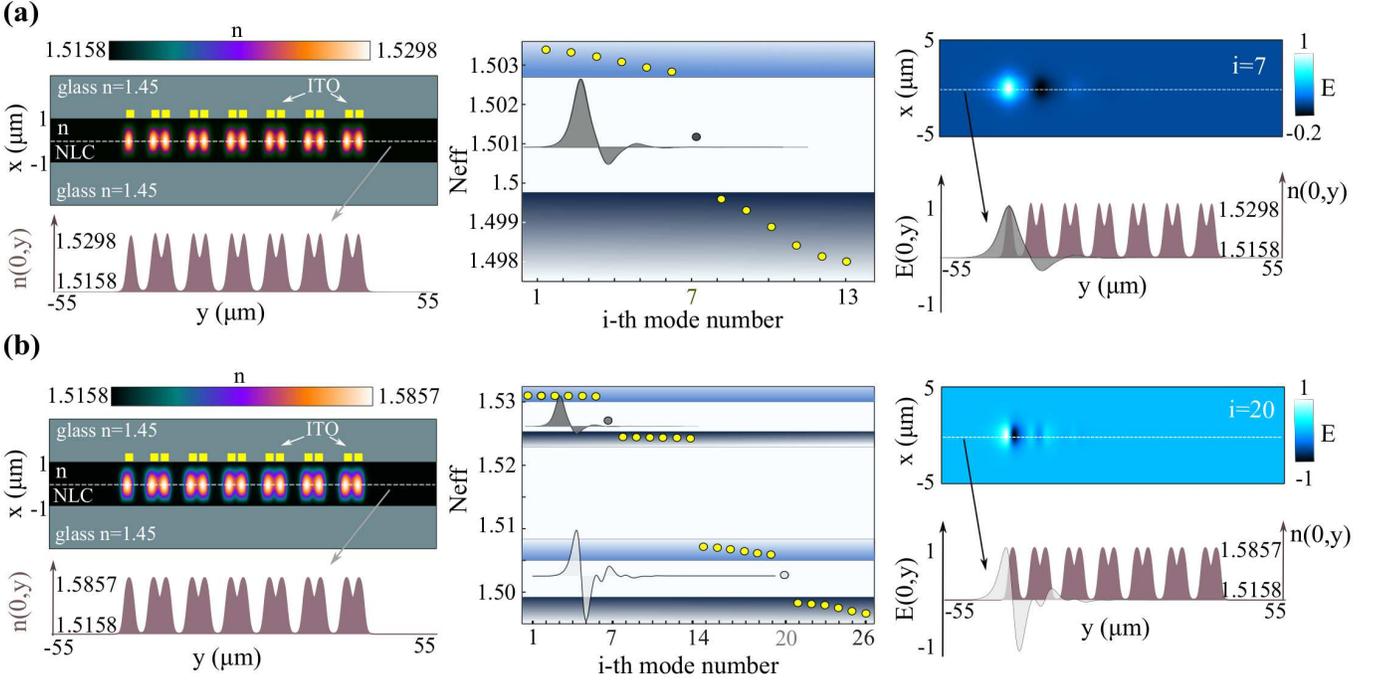}}%
\caption{\label{fig_2}The SSH model in a discrete NLC platform based on thirteen  waveguides operating at $\lambda=1.55\mu m$ induced through the biasing with the voltage of $0.94V$ (a) and  $1.15V$ (b) applied to the electrodes. The left panels show the refractive index distribution in the $xy$ plane. In the middle panel the numerically extracted band structure for two configurations are presented. The right panels in (a) and (b) depict the optical field distribution of the topological states in the $xy$ plane and the cross sections at $x=0$ of the edge modes together with the refractive index distributions. The parameters used in this design are $w=2 \mu m$, $w_d=2.35\mu m$,$d=2 \mu m$, $d_1=6.8\mu m$, $d_2=4 \mu m$, $n_{glass} = 1.45$, NLC $n_o=1.5158$, $n_e=1.6814$, $K=8.15 pN$.}  
\end{figure*} 

For demonstration purposes, let us now consider two examples where the SSH model is implemented in a discrete NLC platform, as depicted in Fig.2(a), (b). In both configurations, the LC cell parameters such as the separation between glass plates $d=2 \mu m$, the electrodes' width  $w=2\mu m$ and $w_d=2.35\mu m$, and the interspace $d_1=6.8 \mu m$ and $d_2=4 \mu m$ are invariant along the propagation direction. At this point, the nonlinear effects are negligible since the arrays operate in the low power regime at a wavelength $\lambda=1.55 \mu m$. After applying external voltage, the electric field from the electrodes reorients the molecules such that the extraordinary refractive index locally increases. As a result, a spatially periodic refractive index distribution is formed as shown in Fig.2(a),(b). The discrete optical lattices consist of 13 waveguides, as shown in  Fig.1(a). The separation distance between the waveguides directly translates into two different coupling strengths $\kappa_{1} $ and $\kappa_{2}$. Here, we consider two cases. In the first scenario, after applying the external voltage ($0.94 V$) to the electrodes each of the waveguides is single-moded. Such structure supports 13 super-modes and a topological edge mode whose eigenvalue is located in the middle of the band-gap (see Fig.2(a)). In the second case, after applying the external voltage ($1.15 V$) each of the waveguides is double-moded. The corresponding spatial refractive index distribution is shown in Fig.2(b). In contrast to the former case, the system has two separate sub-bands and supports in total 26 super-modes and two topological edge modes. The first 13 super-modes including a topological edge mode with the highest eigenvalues are built from the first-order modes of the single waveguide. The successive 13 super-modes with another topological edge mode are constructed from the second-order type states. Again the eigenvalue of these topological states is still located in the middle of each sub-band-gap (Fig.2(b)).

\begin{figure*}[t]
\centerline{\includegraphics[scale=1.0]{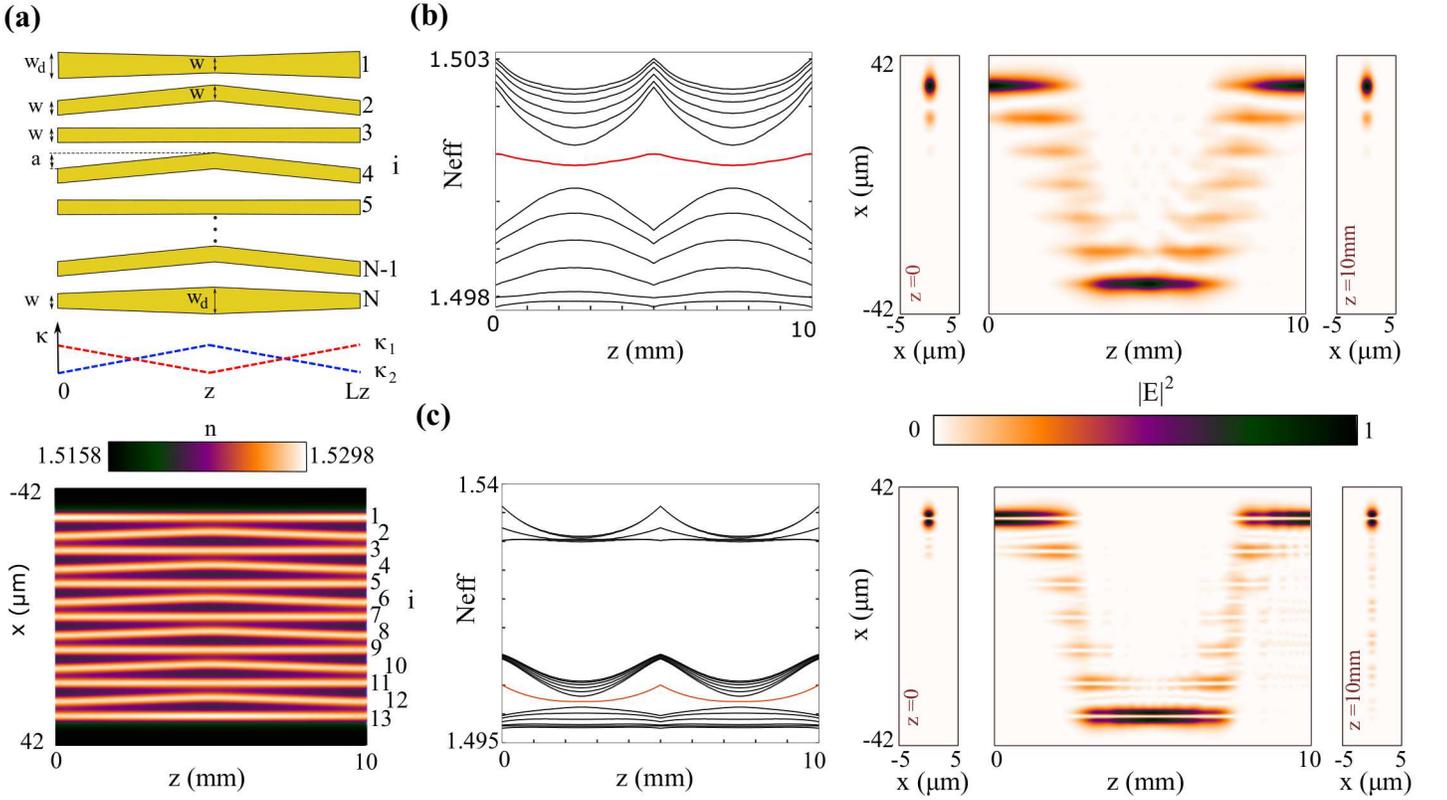}}%
\caption{\label{fig_3}The topological Thouless pumping scheme in a discrete NLS platform. (a) The schematic shape of the ITO electrodes are shown for $w=2\mu m$ and $w_d=2.35\mu m$. Every second electrodes counting from the upper side changes linearly its position along the $z$-direction between $\pm1.4\mu m$ from the central point where the distance between electrodes is equal ($d_1=d_2=5.4 \mu m$). The middle panel shows schematically the varying two couplings coefficients $\kappa_{1}$ and $\kappa_{2}$ in a coupling paths. The bottom panel depicts the slowly varying refractive index distribution in $z$ after applying a finite voltage, forming 13 waveguide arrays. For lower biasing voltage ($0.94V$) these  waveguides are single-moded while they support two modes for higher bias ($1.15V$). (b) The left panel depicts the numerically extracted evolution of the linear spectrum of the system as a function of distance $z$ (a full cycle of the Thuless pump with the topologically protected state is marked in red). The right panels show that the input topological edge mode adiabatically switches between states, during linear evolution, before returning to its initial state at the output when the structure is induced at $0.94V$. (d) The analogical full cycle of the Thouless pump scheme but this time apply for a topological edge mode consisting of the second-order states ($1.15V$).}  
\end{figure*} 

 As we indicated earlier, the topologically protected modes of the 1D SSH array  exhibit no transport properties. The situation radically changes under a topological Thouless pumping protocol.  In our discrete platform, the shape and  orientation of the electrodes are periodically altered  
 along the $z$-direction, as shown in Fig.3(a). After applying the external voltage, the LC molecular orientation establishes again a lattice of  13 waveguides (Fig.3(b)). However, now the  refractive index distribution is varying in both, the $z$ and $y$ direction in order to introduce the required Thouless pumping. In particular, only the position of every second waveguide slowly and periodically varies along the $z$-direction.  The separation between waveguides varies adiabatically,  from $d_1$ to $d_2$ in the first half of the period and from $d_2$ to $d_1$ in the second. In addition, the width of the two outer electrodes vary with distance $z$ to ensure that all local modes of each waveguide share approximately the same propagation constant at a given $z$. At  low optical beam powers (linear regime) such a waveguide lattice  exhibits in $z$ periodically dependent coupling strengths $\kappa_{1}(z)$ and $\kappa_{2}(z)$ as shown schematically in the left panel of Fig. 3(a). The Hamiltonian operator describing the system is now $z$ dependent leading to a $z$-evolution of its spectrum as shown in the left panels in Fig.3(b) and (c) for a complete cycle of the Thouless pump. It is worth noticing that the Hamiltonian operator at the input ($z=0$) is the same as that depicted in Fig.2(a) and (b). The corresponding edge modes (Fig.3(b)(c)) launched into the system now displays transport properties. The light is adiabatically switched from the edge state at $i=1$ at $z=0$ to the edge state that occupies the waveguide number $i=13$  at $z=L_z/2$ and subsequently returns to the initial site $i=1$ at $z=L_z$ . Notice that the transport process does not depend on whether the lattice operate at lowest (Fig.3(c) or higher order mode (Fig.3(d). 
 \begin{figure}[t!]
\centerline{\includegraphics[width=1.0\columnwidth]{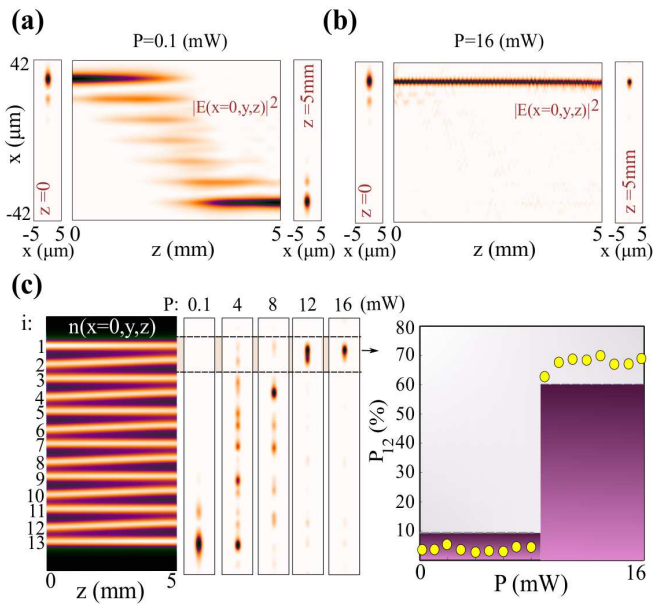}}%
\caption{\label{fig_4} The field evolution in a two-level all-optical switching arrangement that utilizes a topological Thouless pumping scheme with broken chiral symmetry.(a) Evolution of an excited topological edge mode with optical power of 0.1 mW. (b) Nonlinear propagation with the same excitation conditions and input power 16 mW.(c) A threshold all-optical switching based on the ratio between the output power confined in the first two waveguides ($P_{12}$) to the total input power ($P$).}
\end{figure} 
 
 Next, we utilize the topological Thouless pumping scheme under strong nonlinear conditions. To this end, we use the same parameters as in Fig.3(c), but restrict the propagation to the  distance $z=L_z/2=5 mm$. In such configuration, at low powers (0.1 mW), the topologically protected edge state launched at $i=1$ displays transport properties, as depicted in Fig.4(a). Again at the output, a non-trivial edge state forms, with  the energy irreversibly remaining in the waveguide $i=13$. On the other hand, at higher power level (16 mW) and with the same input conditions, the light breaks the chiral symmetry, and propagates in a self-induced nonlinear waveguide (as a discrete soliton). As a result, in the output, we observe most of the energy confined in the first channel ($i=1$). In this respect, all-optical switching occurs between the non-trivial edge state and its trivial nonlinear defect  counterpart. In such configuration, the two level all-optical switching can be observed at approximately $P=9 mW$ of the initial power $P$ as shown in Fig.4(c). For $P\geq 9 mW$, the  system switches to its  upper state with more than $60\%$ of the power confined in the first two waveguides, while below threshold $P<9 mW$, the system occupies the lowest state.

In conclusion, we investigated a nematic liquid crystal SSH topological model  operating under the action of a Thouless pumping scheme. In doing so, we judiciously designed electrode structures to observe a topological edge mode consisting of the superposition of the first and second-order states of the single waveguide. By utilizing a Thouless protocol we observed edge to edge light transport at low power levels. On the other hand, at higher powers,  the transport is frustrated  by light-induced nonlinear defect states, thus giving rise to robust all-optical switching.

This work was partially supported by The  Qatar National Research Fund (grant NPRP13S-0121-200126), the Polish National Science Center (Contract No. UMO-2013/08/A/ST3/00708) and the Polish Ministry of Science and Higher Education (1654/MOB/V/2017/0),the Polish National Science Center (2016/22/M/ST2/00261),the Bodossaki Foundation, DARPA (D18AP00058), the Office of Naval Research (N00014-20-1-2522, N00014-20-1-2789, N00014-16-1-2640, N00014-18-1-2347 and N00014-19-1-2052), FOTECH-1 project ("Electrically-driven waveguiding systems in LC:PDMS structures") granted by the Warsaw University of Technology under the program Excellence Initiative: Research University (ID-UB), the Army Research Office (W911NF-17-1-0481), the Air Force Office of Scientific Research (FA9550-14-1-0037 and FA9550-20-1-0322), the National Science Foundation (CBET 1805200, ECCS 2000538, ECCS 2011171, DMR-1420620), the US–Israel Binational Science Foundation (BSF; 2016381), Simons Foundation (Simons grant 733682).

\nocite{*}

%

\end{document}